\documentclass[english,aps, preprint]{revtex4}
\usepackage[T1]{fontenc}
\usepackage{color}
\usepackage{array}
\usepackage{amsmath}
\usepackage{amssymb}
\usepackage{graphicx}

\makeatletter

\providecommand{\tabularnewline}{\\}

\@ifundefined{textcolor}{}
{%
 \definecolor{BLACK}{gray}{0}
 \definecolor{WHITE}{gray}{1}
 \definecolor{RED}{rgb}{1,0,0}
 \definecolor{GREEN}{rgb}{0,1,0}
 \definecolor{BLUE}{rgb}{0,0,1}
 \definecolor{CYAN}{cmyk}{1,0,0,0}
 \definecolor{MAGENTA}{cmyk}{0,1,0,0}
 \definecolor{YELLOW}{cmyk}{0,0,1,0}
}

\usepackage{times}
\usepackage{hyperref}

\makeatother

\usepackage{babel}
\begin{document}
\title{Electron parallel closures for arbitrary collisionality\\{\small Journal-ref: 
\href{https://dx.doi.org/10.1063/1.4904906}
{\underline{Phys.  Plasmas 21, 112116 (2014)}};
\href{https://dx.doi.org/10.1063/1.4937484}
{\underline{ 22, 129901 (2015)}}}}
\author{Jeong-Young Ji}
\email{j.ji@usu.edu}

\address{Department of Physics, Utah State University, Logan, Utah 84322}
\author{Eric D. Held}
\address{Department of Physics, Utah State University, Logan, Utah 84322}
\begin{abstract}
Electron parallel closures for heat flow, viscosity, and friction
force are expressed as kernel-weighted integrals of thermodynamic
drives, the temperature gradient, relative electron-ion flow velocity,
and flow-velocity gradient. Simple, fitted kernel functions are obtained
for arbitrary collisionality from the 6400 moment solution and the
asymptotic behavior in the collisionless limit. The fitted kernels
circumvent having to solve higher order moment equations in order
to close the electron fluid equations. For this reason, the electron
parallel closures provide a useful and general tool for theoretical
and computational models of astrophysical and laboratory plasmas. 
\end{abstract}
\maketitle

\section{Introduction}

Plasma fluid closures are essential to developing sets of fluid equations
that capture kinetic effects in astrophysical and laboratory plasmas.
A complete set of closures for the Maxwellian moment equations (for
density $n$, temperature $T$, and flow velocity $\mathbf{V}$) exists
only for high collisionality~\citep{Braginskii1965}. At low collisionality,
particle free streaming parallel to the magnetic field becomes significant
and closures are affected by thermodynamic drives along the field
line. Consequently, the closures take on an integral form, as they
do in the collisionless limit~\citep{Hammett1990P,Hazeltine1998}. 

Obtaining quantitative parallel closures for arbitrary collisionality
requires accurate collision operators. As a crude approximation, a
Krook type operator was adopted in the 3+1 closure model of Ref.~\citep{Snyder1997HD}.
As better approximations, Lorentz type operators were used to obtain
closures in wave number space~\citep{Chang1992C} and approximate
integral closures for the heat flow due to a temperature gradient~\citep{Held2001CHS}
and viscosity due to a flow velocity gradient~\citep{Held2003}. 

Recently, the exact linearized Coulomb collision operators were analytically
calculated for an infinite hierarchy of moment equations~\citep{Ji2006H,Ji2008H}.
These moment equations were truncated, linearized, and solved to produce
integral parallel closures for arbitrary collisionality~\citep{Ji2009HS,Ji2009H1}.
The resultant integral heat flow closure was tested in numerical calculations
of energy confinement in the Sustained Spheromak Physics Experiment
and yielded modestly better agreement with experimental measurements
than Braginskii's diffusive heat flow~\citep{Ji2009HS}. This integral
heat flow was also used to capture kinetic effects in the JET scrape-off
layer~\citep{Omotani2013D}. 

Practical use of the integral closures is complicated by the need
to express the kernel functions for arbitrary collisionality. Kernels
obtained from $N$-moment equations are sums of $N/2$ exponential
functions. At low collisionality, large $N$ is needed for convergent
results. In the collisionless limit, however, simple analytical expressions
for the kernels exist~\citep{Hammett1990P,Chang1992C,Hazeltine1998,Ji2013HJ}.
In this paper, we provide simple kernel functions that map onto the
$N=6400$ kernels in the convergent regime and the collisionless kernels
in the nonconvergent regime. 

\section{Parallel moment equations and closures}

For a strong magnetic field, parallel moment equations can be obtained
by taking moments of the drift kinetic equation~\citep{Hazeltine1973}
or, equivalently, by taking the parallel component of general moment
equations~\citep{Ji2014H}. For no magnetic field, an inhomogeneous
system along one direction produces similar moment equations. The
linearized parallel moment equations are~\citep{Ji2009HS,Ji2009H1}
\begin{equation}
\sum_{B=1}^{N}\Psi_{AB}\frac{\partial n_{B}}{\partial\eta}=\sum_{B=1}^{N}C_{AB}n_{B}+g_{A},\label{MEll}
\end{equation}
where the moment index $A=1,2,\cdots,N$ enumerates $(l,k)=(0,2),\cdots,(0,K+1),(1,1),\cdots,(1,K),(2,0),\cdots,(L-1,K-1)$.
The indices $l$ and $k$ denote the Legendre polynomial and Laguerre-Sonine
polynomial orders with $N=LK$. The arclength $\ell$ along a magnetic
field line is normalized by the mean free path, $d\eta=d\ell/\lambda_{\mathrm{mfp}}$
or $\eta(\ell)=\int^{\ell}d\ell^{\prime}/\lambda_{\mathrm{mfp}}(\ell^{\prime})$,
where $\lambda_{\mathrm{mfp}}=v_{T}\tau$, the electron thermal speed
$v_{T}=\sqrt{2T/m}$, and the electron-electron collision time $\tau=\tau_{\mathrm{ee}}=6\sqrt{2}\pi^{3/2}\epsilon_{0}^{2}\sqrt{m}T^{3/2}/ne^{4}\ln\Lambda$.
Here $m$ is the electron mass, $e$ is the electron charge, and $\ln\Lambda$
is the Coulomb logarithm. The collision matrix $C_{AB}$ includes
the electron-electron and electron-ion collision terms. For electrons,
the nonvanishing thermodynamic drives $g_{A}$ include the parallel
temperature gradient $\partial_{\|}T$, the relative parallel flow
velocity $V_{\mathrm{ei}\|}$, and the parallel component of the rate
of strain tensor $W_{\|}=\mathbf{b}\mathbf{b}:\mathsf{W},\;(\mathsf{W})_{\alpha\beta}=\partial_{\alpha}V_{\beta}+\partial_{\beta}V_{\alpha}-\frac{2}{3}\delta_{\alpha\beta}\nabla\cdot\mathbf{V}$.
The non-Maxwellian moments $n_{A}$ include the parallel heat flux
density $h_{\|}$, parallel viscosity $\pi_{\|}$, and parallel friction
force density $R_{\|}$.

The linear system \eqref{MEll} with constant matrices $\Psi$ and
$C$ is solved by computing the eigensystem of $\Psi^{-1}C$~\citep{Ji2009HS,Ji2009H1}.
The desired particular solution is 
\begin{eqnarray}
h_{\|}(\ell) & = & Tv_{T}\int d\eta^{\prime}\Bigl(-\frac{1}{2}K_{hh}\frac{n}{T}\frac{dT}{d\eta^{\prime}}+K_{hR}Zn\frac{V_{\mathrm{ei}\|}}{v_{T}}{\color{red}-}K_{h\pi}\frac{3}{4}n\tau W_{\|}\Bigr),\label{h:}\\
R_{\|}(\ell) & = & -\frac{mn}{\tau_{\mathrm{ei}}}V_{\mathrm{ei}\|}+\frac{mv_{T}}{\tau_{\mathrm{ei}}}\int d\eta^{\prime}\Bigl(-K_{Rh}\frac{n}{2T}\frac{dT}{d\eta^{\prime}}+K_{RR}Zn\frac{V_{\mathrm{ei}\|}}{v_{T}}{\color{red}-}K_{R\pi}\frac{3}{4}n\tau W_{\|}\Bigr),\label{R:}\\
\pi_{\|}(\ell) & = & T\int d\eta^{\prime}\Bigl({\color{red}-}K_{\pi h}\frac{n}{T}\frac{dT}{d\eta^{\prime}}{\color{red}+}2K_{\pi R}Zn\frac{V_{\mathrm{ei}\|}}{v_{T}}-K_{\pi\pi}\frac{3}{4}n\tau W_{\|}\Bigr),\label{p:}
\end{eqnarray}
where $\int d\eta^{\prime}K_{AB}g_{B}$ means $\int d\eta^{\prime}K_{AB}(\eta-\eta^{\prime})g_{B}(\ell^{\prime})$
with $\eta=\eta(\ell)$ and $\eta^{\prime}=\eta(\ell^{\prime})$.
The $N$-moment kernels are given by 
\begin{equation}
K_{AD}(\eta)=\begin{cases}
{\displaystyle -\sum_{\{B|k_{B}>0\}}^{N}\gamma_{AD}^{B}e^{{\color{red}+}k_{B}\eta}}, & \eta{\color{red}<}0,\\
{\displaystyle +\sum_{\{B|k_{B}<0\}}^{N}\gamma_{AD}^{B}e^{{\color{red}+}k_{B}\eta}}, & \eta{\color{red}>}0,
\end{cases}\label{K:e}
\end{equation}
where eigenvalues $k_{B}$ of $\Psi^{-1}C$ appear in positive and
negative pairs. The coefficients $\gamma_{AD}^{B}$ are determined
by the eigenvector components and satisfy $\gamma_{AD}^{B}=\gamma_{DA}^{B}$
and 
\begin{equation}
\gamma_{AD}^{-B}=\begin{cases}
-\gamma_{AD}^{B}, & AD=hh,hR,RR,\pi\pi\equiv\mathrm{even},\\
+\gamma_{AD}^{B}, & AD=h\pi,R\pi\equiv\mathrm{odd},
\end{cases}\label{gam+-}
\end{equation}
where $-B$ denotes the moment index corresponding to $-k_{B}$. Therefore
$K_{AD}=K_{DA}$ and the kernels are even or odd functions of $\eta$.

\begin{figure}
\includegraphics[clip,scale=0.95]{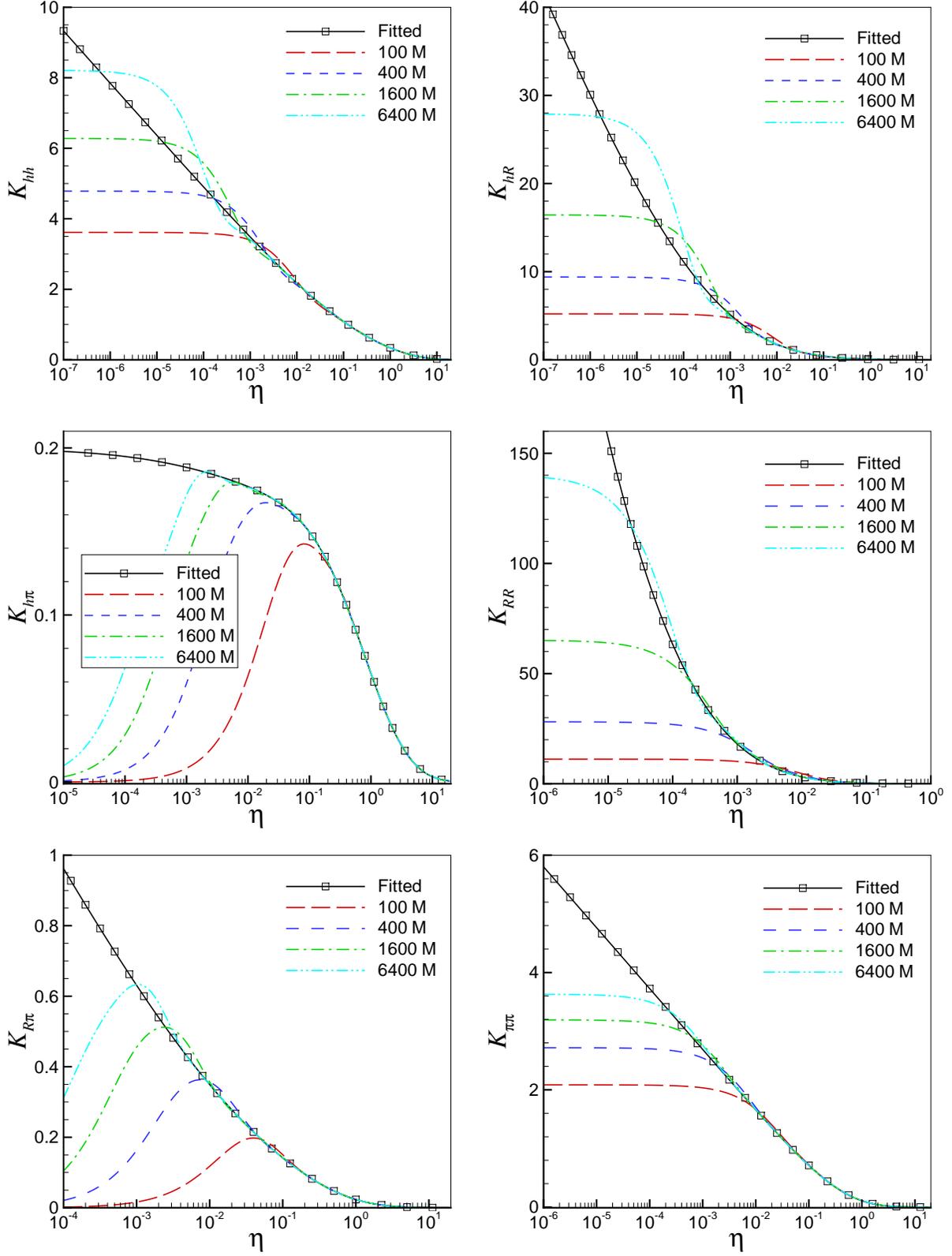}\caption{(Color online) Kernel functions for ion charge $Z=1$ computed from
$N=$100 (red, long-dashed), 400 (blue, short-dashed), 1600 (green,
dash-dotted), and 6400 (cyan, dash-dot-dotted). The fitted kernels
(black, solid with squares) are also shown.}

\label{fig:K}
\end{figure}

Since the kernels are computed from a truncated system of $N$ moment
equations, convergence should be checked as $N$ increases. We start
with $N=100$ ($L=10$ and $K=10$) and go to $N=6400$ ($L=80$ and
$K=80$) by quadrupling $N$ (doubling both $L$ and $K$). The convergence
of the kernel functions is shown in Fig.~\ref{fig:K}.
 In the collisional limit, the drives are effectively constant and
the areas $\int K_{AB}(\eta)d\eta$ become the classical closure coefficients.
The areas change little with increasing $N$, indicating that the
$N=100\ (K=10)$ moment calculation accurately yields the classical
coefficients. In general, the kernels converge for $\eta\gtrsim\eta_{\mathrm{c}}$,
where $\eta_{\mathrm{c}}$ decreases with increasing $N$. 

To study the convergence of closures as collisionality varies, we
consider sinusoidal drives, $T=T_{0}+T_{1}\sin\varphi$, $V_{\|}=V_{0}+V_{1}\sin\varphi$,
and $V_{\mathrm{ei}\|}=V_{\mathrm{ei}}\cos\varphi$, where $\varphi=2\pi\ell/\lambda+\varphi_{0}=k\eta+\varphi_{0}$.
The wave number normalized by the mean free path, $k=2\pi\lambda_{\mathrm{mfp}}/\lambda$,
measures the inverse collisionality. Note that although $\lambda$
and $\varphi_{0}$ (and hence $\varphi$) may be different for each
drive, we have used the same symbols for convenience. Assuming that
$n$ and $v_{T}\approx\sqrt{2T_{0}/m}$ are constant and that $\mathbf{V}_{\perp}=0$,
for simplicity, the closures can be obtained by evaluating the integrals,
\begin{equation}
\int K_{AD}(\eta-\eta^{\prime})\cos\varphi^{\prime}d\eta^{\prime}=\begin{cases}
\hat{K}_{AD}\cos\varphi, & AD=\mbox{even},\\
{\displaystyle \hat{K}}_{AD}\sin\varphi, & AD=\mbox{odd},
\end{cases}
\end{equation}
where $\varphi^{\prime}=k\eta^{\prime}+\varphi_{0}$ and
\begin{equation}
\hat{K}_{AD}=\begin{cases}
{\displaystyle \sum_{B=1}^{N}\frac{-\gamma_{AD}^{B}k_{B}}{k_{B}^{2}+k^{2}}}, & AD=\mbox{even},\\
{\displaystyle \sum_{B=1}^{N}\frac{\gamma_{AD}^{B}k}{k_{B}^{2}+k^{2}}}, & AD=\mbox{odd}.
\end{cases}
\end{equation}
Upon linearization the closures become
\begin{eqnarray}
h_{\|}(\ell) & = & -\frac{1}{2}nT_{1}v_{T}\hat{h}_{h}\cos\varphi+nT_{0}ZV_{\mathrm{ei}}\hat{h}_{R}\cos\varphi{\color{red}-}nT_{0}V_{1}\hat{h}_{\pi}\sin\varphi,\\
R_{\|}(\ell) & = & -nT_{1}\frac{2\pi}{\lambda}\hat{R}_{h}\cos\varphi-\frac{mnV_{\mathrm{ei}}}{\tau_{\mathrm{ei}}}\hat{R}_{R}\cos\varphi{\color{red}-}nmV_{1}\frac{2\pi v_{T}}{\lambda}\hat{R}_{\pi}\sin\varphi,\\
\pi_{\|}(\ell) & = & {\color{red}-}nT_{1}\hat{\pi}_{h}\sin\varphi{\color{red}+}2nT_{{\color{red}0}}Z\frac{V_{\mathrm{ei}}}{v_{T}}\hat{\pi}_{R}\sin\varphi-nT_{0}\frac{V_{1}}{v_{T}}\hat{\pi}_{\pi}\cos\varphi,
\end{eqnarray}
with the dimensionless closures defined by $\hat{h}_{h}=k\hat{K}_{hh},$
$\hat{h}_{R}={\color{red}Z}\hat{K}_{hR}=\hat{R}_{h},$ $\hat{h}_{\pi}=k\hat{K}_{h\pi}=\hat{\pi}_{h},$
$\hat{R}_{R}=1-\hat{K}_{RR},$ $\hat{R}_{\pi}={\color{red}Z}\hat{K}_{R\pi}=\hat{\pi}_{R},$
and $\hat{\pi}_{\pi}=k\hat{K}_{\pi\pi}.$ 

\begin{figure}
\includegraphics[clip,scale=0.95]{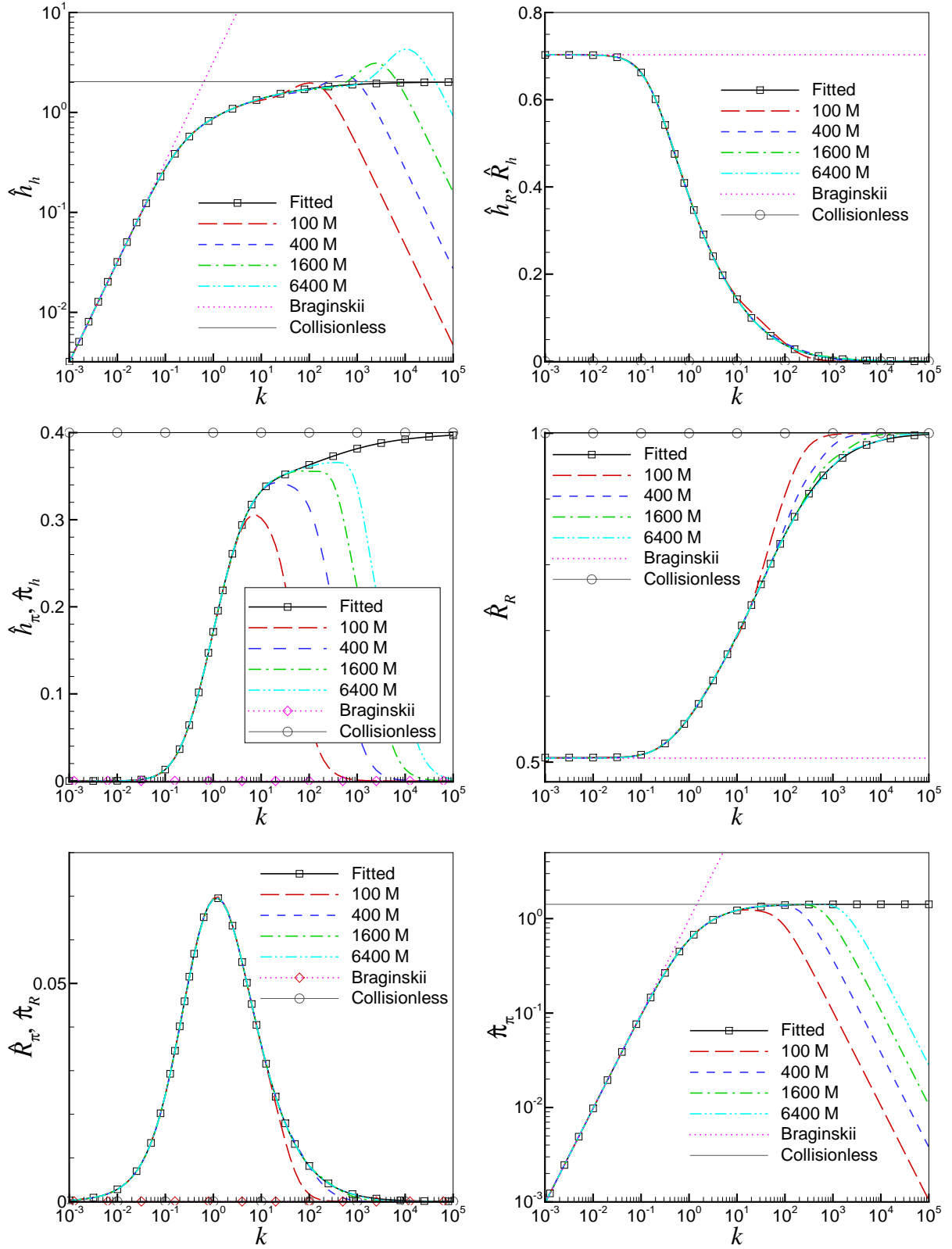}\caption{(Color online) Closures for sinusoidal drives computed for various
$N$ and fitted kernels in Fig.~\ref{fig:K}. The Braginskii (red,
dotted {[}with diamonds{]}) and collisionless (gray, thin solid {[}with
squares{]}) closures are also shown.}
\label{fig:n}
\end{figure}

Figure \ref{fig:n} shows the dimensionless closures as functions
of $k$. For a given gradient scale length $\lambda$, as $\lambda_{\mathrm{mfp}}$
increases (collisionality decreases), $k$ increases. For small $k$
(high collisionality), the closures are convergent with a small number
of moments (see Fig.~\ref{fig:n}). 
 Errors in Braginskii's closures are largest for $h_{h}$, the heat
flow due to the temperature gradient. This error is 16\% for $k\sim0.1\;(\epsilon\equiv\lambda_{\mathrm{mfp}}/\lambda\sim0.02)$
and 100\% for $k\sim0.4\;(\epsilon\sim0.06)$. These results agree
well with the prediction of the higher-order Chapman-Enskog theory
in Ref.~\citep{Ji2006H}. In Figs.~\ref{fig:K} and \ref{fig:n},
the convergence of closures is attributed to the convergence of the
kernels for $\eta\gtrsim\eta_{\mathrm{c}}$. Note that $\eta_{\mathrm{c}}$
decreases as $N$ increases. As $k$ increases, convergent closures
are obtained by using more moments. Although, for a given $k$, there
exists $N$ that yields convergent results, it is more profitable
to invoke the collisionless theory. Instead of increasing $N$ (decreasing
$\eta_{\mathrm{c}}$) to obtain accurate kernels for smaller $\eta$,
we consider mapping onto the collisionless kernels for $\eta\ll1$
in order to determine fitted kernels that are approximately valid
for arbitrary collisionality.

\section{Simple fitted kernel functions}

The kernel functions obtained from 6400 moment equations include 3200
terms {[}see Eq.~\eqref{K:e}{]}. Incorporating those eigenvalues
and the corresponding coefficients in simulations is impractical.
Furthermore, the kernels are still inaccurate for $k\gtrsim80$ near
the collisionless limit. Therefore, it is convenient to find simple
fitted functions that (i) accurately represent the moment solutions
in the convergent regime $(k\lesssim80)$, and (ii) have the correct
asymptotic behavior in the collisionless limit. This approach yields
kernels that are nearly exact for arbitrary collisionality.

In the collisional limit $(\lambda_{\mathrm{mfp}}\rightarrow0)$,
the parallel closures are described by Braginskii's theory ($K=2$
calculation~\citep{Braginskii1965}, here convergent values~\citep{Ji2013H}
are shown) with ion charge $Z=1$, 
\begin{eqnarray}
h_{\|} & = & -3.203\frac{nT\tau_{\mathrm{ee}}}{m}\partial_{\|}T+0.703nTV_{\mathrm{ei}\|},\label{hc:}\\
R_{\|} & = & -0.703n\partial_{\|}T-0.506\frac{mn}{\tau_{\mathrm{ei}}}V_{\mathrm{ei}\|},\label{Rc:}\\
\pi_{\|} & = & -0.978nT\tau_{\mathrm{ee}}\frac{3}{4}W_{\|}.\label{pc:}
\end{eqnarray}
Noting that $\int d\eta^{\prime}K_{AD}(\eta-\eta^{\prime})g_{D}(\ell^{\prime})$$\rightarrow\hat{\kappa}_{AD}g_{D}(\ell)$
in the collisional limit, the fitted kernels should satisfy 
\begin{equation}
\int_{-\infty}^{\infty}d\eta\left(\begin{array}{c}
K_{hh}(\eta)\\
K_{hR}(\eta)\\
K_{RR}(\eta)\\
K_{\pi\pi}(\eta)
\end{array}\right)=\left(\begin{array}{c}
3.203\\
0.703\\
0.494\\
0.978
\end{array}\right).\label{aK}
\end{equation}
The coefficients connecting $h_{\|}$ ($\pi_{\|})$ to $W_{\|}$ ($\partial_{\|}T$)
and $R_{\|}$ ($\pi_{\|})$ to $W_{\|}$ ($V_{\mathrm{ei}\|}$) in
Eqs.~\eqref{hc:}-\eqref{pc:} vanish because the corresponding kernels
are odd functions. However, we evaluate the integrals over $[0,\infty)$
to make the fitted kernels satisfy 
\begin{equation}
\int_{0}^{\infty}d\eta\left(\begin{array}{c}
K_{h\pi}(\eta)\\
K_{R\pi}(\eta)
\end{array}\right)=\left(\begin{array}{c}
0.264\\
0.104
\end{array}\right).\label{aK1}
\end{equation}

Finally, the asymptotic behavior of the kernels for small $\eta$
can be obtained from closures in the collisionless limit~\citep{Ji2013HJ}.
For $\eta\ll1$, we have 
\begin{eqnarray}
K_{hh}(\eta) & \approx & -\frac{18}{5\pi^{3/2}}(\ln|\eta|+\gamma_{h}),\nonumber \\
K_{hp}(\eta) & \approx & \frac{1}{5},\\
K_{pp}(\eta) & \approx & -\frac{4}{5\pi^{1/2}}(\ln|\eta|+\gamma_{p}),\nonumber 
\end{eqnarray}
where $\gamma_{h}$ and $\gamma_{p}$ are constants. For $K_{hR}$,
$K_{RR}$, and $K_{R\pi}$ (friction related kernels), the asymptotic
forms do not exist. However, the corresponding closures vanish in
the collisionless limit and extrapolations of the $6400$-moment kernels
are accurate enough. 

Putting all of this together, the kernel functions can be fitted to
a single function 
\begin{equation}
K_{AB}(\eta)=-[d+a\exp(-b\eta^{c})]\ln[1-\alpha\exp(-\beta\eta^{\gamma})]\label{Kfit}
\end{equation}
with parameters \emph{a, b, c, d,} $\alpha$, $\beta$, and $\gamma$
listed in Table \ref{t:K}. 
\begin{table}
\begin{tabular}{>{\centering}p{0.12\textwidth}>{\centering}p{0.12\textwidth}>{\centering}p{0.12\textwidth}>{\centering}p{0.12\textwidth}>{\centering}p{0.12\textwidth}>{\centering}p{0.12\textwidth}>{\centering}p{0.12\textwidth}>{\centering}p{0.12\textwidth}}
\hline 
 & $a$ & $b$ & $c$ & $d$ & $\alpha$ & $\beta$ & $\gamma$\tabularnewline
\hline 
$K_{hh}$ & -5.32 & 0.170 & 0.646 & 6.87 & 1 & 2.02 & 0.417\tabularnewline
$K_{hR}$ & 6.37 & 5.12 & 0.160 & 0.100 & 1 & 1 & 0.583\tabularnewline
$K_{h\pi}$ & -0.229 & 2.26 & 0.594 & 0.363 & 0.775 & 1.49 & 0.478\tabularnewline
$K_{RR}$ & 245 & 8.06 & 0.147 & 0.432 & 1 & 3.40 & 0.347\tabularnewline
$K_{R\pi}$ & -0.226 & 3.21 & 0.678 & 0.696 & 1 & 3.40 & 0.347\tabularnewline
$K_{\pi\pi}$ & 0.724 & 0.932 & 0.654 & 0.195 & 1 & 1.60 & 0.491\tabularnewline
\hline 
\end{tabular}\caption{Fitted parameters for kernel function $K_{AB}(\eta)=-[d+a\exp(-b\eta^{c})]\ln[1-\alpha\exp(-\beta\eta^{\gamma})]$.}
\label{t:K}
\end{table}
 The fitted kernels are plotted in Fig.~\ref{fig:K} and the corresponding
closure moments due to sinusoidal drives in Fig.~\ref{fig:n}. Note
that the fitted kernels reproduce closure values in the collisional
$(k\rightarrow0)$ and collisionless $(k\rightarrow\infty)$ limits. 

\section{Discussion}

We have presented a complete set of electron parallel closures with
ion charge $Z=1$ for arbitrary collisionality, Eqs.~\eqref{h:}-\eqref{p:}
and Table \ref{t:K}. Generalization to arbitrary $Z$ is straightforward.
Ion parallel closures can also be obtained similarly for varying electron-ion
temperature ratios. Efforts to obtain analytic expressions for these
cases are ongoing and will be presented in the near future.

Closing fluid equations using the integral closures presented here
is computationally feasible in simulations of astrophysical and laboratory
plasmas. Although the implementation of such closures benefits from
field-line aligned coordinates (BOUT++~\citep{Dudson2009e4}), plasma
fluid codes that do not have field-line aligned grids (e.g., NIMROD~\citep{Sovinec2004-9})
may still benefit from the added efficiency of evaluating simple fitted
kernels. The practicability of three dimensional fluid simulations
with integral closures is shown in Ref.~\citep{Ji2009HS} and discussed
in Ref.~\citep{Omotani2013D}. 

\section*{Acknowledgments}

One of authors (Ji) would like to thank Dr. Hogun Jhang for general
discussions of using the integral closures and Dr. Dongcheol Seo for
discussing the fitting method. The research was supported by the U.S.
DOE under grant Nos. DE-FG02-04ER54746, DE-FC02-04ER54798 and DE-FC02-05ER54812
and by the World Class Institute (WCI) Program of the National Research
Foundation of Korea (NRF) funded by the Ministry of Education, Science
and Technology of Korea (MEST) (NRF Grant No. WCI 2009-001). This
work is performed in conjunction with the Plasma Science and Innovation
(PSI) center and the Center for Extended Magnetohydrodynamics. This
research used resources of the National Energy Research Scientific
Computing Center, which is supported by the Office of Science of the
U.S. Department of Energy under Contract No. DE-AC02-05CH11231.

\end{document}